\documentclass{desyproc}

\newcommand{\lwig}{\mbox{\;\raisebox{.3ex}
    {$<$}$\!\!\!\!\!$\raisebox{-.9ex}{$\sim$}\;}}

\begin{document}

\title{Axion hot dark matter bounds}

\author{{\slshape G.~Raffelt$\,^1$, S.~Hannestad$\,^2$
A.~Mirizzi$\,^1$, Y.~Y.~Y.~Wong$\,^1$}\\[1ex]
$^1$Max-Planck-Institut f\"ur Physik, F\"ohringer Ring 6,
D-80805 M\"unchen, Germany\\
$^2$Department of Physics and Astronomy,
University of Aarhus, DK-8000 Aarhus C, Denmark}

\contribID{raffelt\_georg}
\desyproc{DESY-PROC-2008-02}
\acronym{Patras 2008}
\doi

\maketitle

\begin{abstract}
We derive cosmological limits on two-component hot dark matter
consisting of neutrinos and axions. We restrict the large-scale
structure data to the safely linear regime, excluding the
Lyman-$\alpha$ forest. We derive Bayesian credible regions in the
two-parameter space consisting of $m_a$ and $\sum m_\nu$.
Marginalizing over $\sum m_\nu$ provides $m_a< 1.02$~eV (95\% C.L.).
In the absence of axions the same data and methods give $\sum m_\nu<
0.63$~eV (95\% C.L.).
\end{abstract}

\section{Introduction}

The masses of the lightest particles are best constrained by the
largest cosmic structures. The well-established method of using
cosmological precision data to constrain the cosmic hot dark matter
fraction~\cite{Lesgourgues:2006nd, Hannestad:2006zg} has been extended
to hypothetical low-mass particles, notably to axions, in several
papers~\cite{Hannestad:2003ye, Hannestad:2005df, Melchiorri:2007cd,
  Hannestad:2007dd, Hannestad:2008js}. If axions thermalize after the
QCD phase transition, their number density is comparable to that of
one neutrino family. Neutrino mass limits are in the sub-eV range so
that axion mass limits will be similar and therefore of interest to
experiments like CAST~\cite{Andriamonje:2007ew} or the Tokyo axion
helioscope~\cite{Inoue:2008zp} that search for axions in the mass
range around 1~eV. We here summarize our detailed limits on axions
that were derived from the latest sets of cosmological data, including
WMAP (5~years). Numerically our latest limit on
$m_a$~\cite{Hannestad:2008js} is almost identical to one that some of
us have derived several years ago~\cite{Hannestad:2005df}. The main
difference is that in the older paper the Lyman-$\alpha$ data were
used that we now consider ``too dangerous'' in that they are prone to
systematic errors. So, the latest data, that are safely in the linear
regime, now do as well as the older data where Lyman-$\alpha$ was
included.

\section{Axions}

The Peccei-Quinn solution of the CP problem of strong interactions
predicts the existence of axions, low-mass pseudoscalars that are
very similar to neutral pions, except that their mass and
interaction strengths are suppressed by a factor of order
$f_\pi/f_a$, where $f_\pi\approx 93$~MeV is the pion decay constant,
and $f_a$ the axion decay constant or
Peccei--Quinn scale~\cite{Peccei:2006as}. In more detail, the axion
mass~is
\begin{equation}\label{eq:axmass}
 m_a=\frac{z^{1/2}}{1+z}\,\frac{f_\pi m_\pi}{f_a}
 =\frac{6.0~{\rm eV}}{f_a/10^6~{\rm GeV}}\,,
\end{equation}
where $z=m_u/m_d$ is the mass ratio of up and down quarks.  A value
$z=0.56$ was often assumed, but it could vary in the range 0.3--0.6
\cite{Yao:2006px}.  A large range of $f_a$ values is excluded by
experiments and by astrophysical and cosmological
arguments~\cite{Raffelt:2006cw}. Axions with a mass of order
10~$\mu$eV could well be the cold dark matter of the
universe~\cite{Sikivie:2006ni} and if so will be found eventually by
the ongoing ADMX experiment, provided that $1~\mu{\rm
  eV}<m_a<100~\mu{\rm eV}$~\cite{Asztalos:2006kz}.

\begin{figure}[h]
\centerline{\includegraphics[width=0.7\textwidth]%
{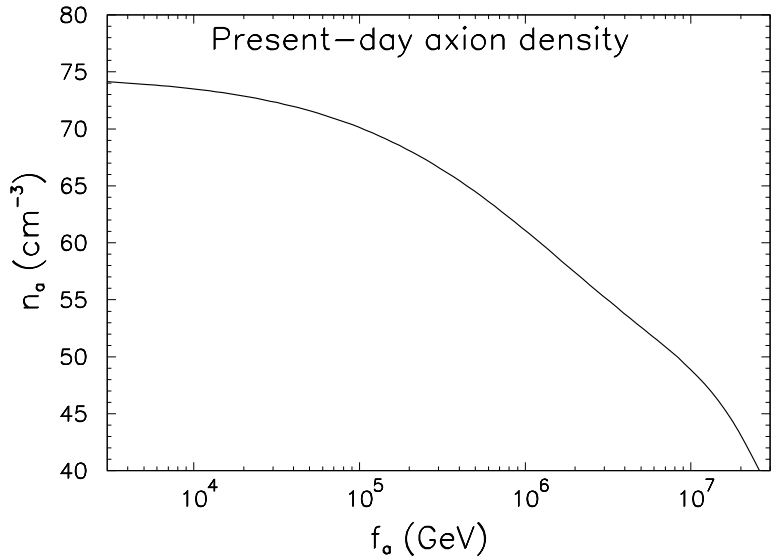}}
\caption{Axion relic density.\label{fig:relic}}
\end{figure}

In addition, a hot axion population is produced by thermal
processes~\cite{Chang:1993gm, Masso:2002np}. Axions attain thermal
equilibrium at the QCD phase transition or later if
$f_a\lwig10^8$~GeV, erasing the cold axion population produced
earlier and providing a hot dark matter component instead.  If
axions do not couple to charged leptons (``hadronic axions'') the
main thermalization process in the post-QCD epoch
is~\cite{Chang:1993gm} $a+\pi\leftrightarrow\pi+\pi$.  The
axion--pion interaction is given by a Lagrangian of the
form~\cite{Chang:1993gm} ${\cal
  L}_{a\pi}=(C_{a\pi}/f_\pi f_a)\, (\pi^0\pi^+\partial_\mu\pi^-
+\pi^0\pi^-\partial_\mu\pi^+ -2\pi^+\pi^-\partial_\mu\pi^0)
\partial_\mu a$.  In hadronic axion models, the coupling constant
is~\cite{Chang:1993gm} $C_{a\pi}=(1-z)/[3\,(1+z)]$.  Based on this
interaction, the axion decoupling temperature in the early universe
was calculated in Ref.~\cite{Hannestad:2005df}, where all relevant
details are reported. In Fig.~\ref{fig:relic} we show the relic axion
density as a function of $f_a$.

\section{Cosmological model and data}

We consider a cosmological model with vanishing spatial curvature and
adiabatic initial conditions, described by six free parameters, the
dark-matter density $\omega_{\rm dm}=\Omega_{\rm dm} h^2$, the baryon
density $\omega_b=\Omega_b h^2$, the Hubble parameter $H_0=h~100~{\rm
  km~s^{-1}~Mpc^{-1}}$, the optical depth to reionization $\tau$, the
amplitude of the primordial scalar power spectrum $\ln(10^{10}A_s)$,
and its spectral index $n_s$. In addition we allow for a nonzero sum
of neutrino masses $\sum m_\nu$ and a nonvanishing axion mass $m_a$
which also determines the relic density shown in Fig.~\ref{fig:relic}
by the standard relation between $m_a$ and $f_a$. We show the priors
on our parameters in Ref.~\cite{Hannestad:2007dd}.

We use the 5-year release of the WMAP cosmic microwave
data~\cite{Hinshaw:2008kr, Nolta:2008ih} that we analyze using
version~3 of the likelihood calculation package provided by the WMAP
team on the LAMBDA homepage~\cite{lambda}, following closely the
analyses of references~\cite{Dunkley:2008ie, Komatsu:2008hk}.  For
the large-scale galaxy power spectra we use $P_{\rm g}(k)$ inferred
from the luminous red galaxy (LRG) sample of the Sloan Digital Sky
Survey (SDSS) \cite{Percival:2006gt, Tegmark:2006az} and from the
Two-degree Field Galaxy Redshift Survey (2dF)~\cite{Cole:2005sx}. We
only use data safely in the linear regime where a scale-independent
bias is likely to hold true. For 2dF this is $k_{\rm max} \sim 0.09
\ h \ {\rm Mpc}^{-1}$ (17 bands) and for SDSS-LRG $k_{\rm max} \sim
0.07 \ h \ {\rm Mpc}^{-1} $ (11 bands). We do not use Lyman-$\alpha$
data at all.  The baryon acoustic oscillation peak was measured in
the SDSS luminous red galaxy sample~\cite{Eisenstein2005}.  We use
all 20~points in the two-point correlation data and the
corresponding analysis procedure~\cite{Eisenstein2005}. We use the
SN~Ia luminosity distance measurements of provided by Davis
et~al.~\cite{Davis:2007na}.

\section{Results}

\begin{figure}[hb]
\centerline{\includegraphics[width=0.6\textwidth]{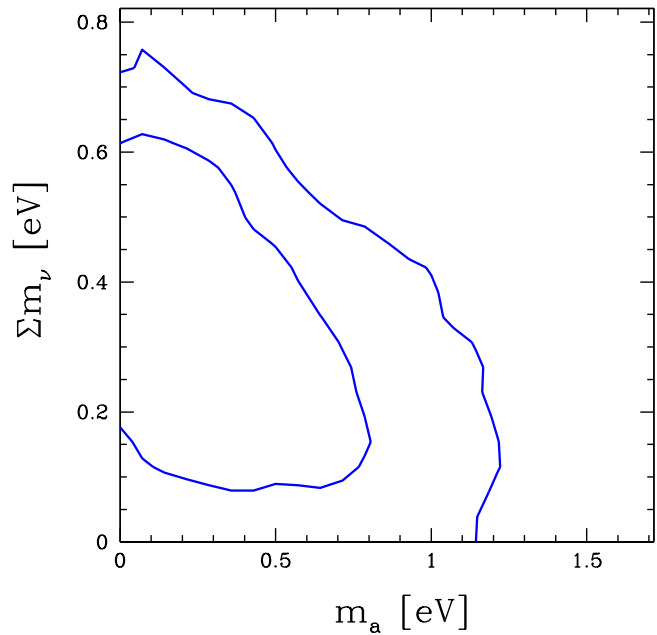}}
\caption{2D marginal 68\% and 95\% contours in the $\sum m_\nu$--$m_a$
  plane.\label{fig:bounds}}
\end{figure}

We use standard Bayesian inference techniques and explore the model
parameter space with Monte Carlo Markov Chains (MCMC) generated using
the publicly available {\sc CosmoMC} package~\cite{Lewis:2002ah,
  cosmomc}. We find the 68\% and 95\% 2D marginal contours shown in
Fig.~\ref{fig:bounds} in the parameter plane of $\sum m_\nu$ and
$m_a$.  Marginalizing over $\sum m_\nu$ provides $m_a< 1.02$~eV (95\%
C.L.).  In the absence of axions the same data and methods give $\sum
m_\nu< 0.63$~eV (95\% C.L.).  These axion mass limits are nicely
complementary to the search range of the CAST
experiment~\cite{Andriamonje:2007ew} and the Tokyo
helioscope~\cite{Inoue:2008zp} that can reach to 1~eV or somewhat
above. While the hot dark matter limits are not competitive with the
SN~1987A limits, it is intriguing that cosmology alone now
provides both an upper and a lower limit for the allowed range of
axion parameters.

\section*{Acknowledgements}

We acknowledge use of computing resources from the Danish Center for
Scientific Computing (DCSC) and partial support by the European Union
under the ILIAS project (contract No.\ RII3-CT-2004-506222), by the
Deutsche Forschungsgemeinschaft under the grant TR-27 ``Neutrinos and
Beyond'' and by The Cluster of Excellence for Fundamental Physics
``Origin and Structure of the Universe'' (Garching and Munich).

\begin{footnotesize}

\end{footnotesize}

\begin{thebibliography}{99}

\bibitem{Lesgourgues:2006nd}
  J.~Lesgourgues and S.~Pastor,
  Phys.\ Rept.\  {\bf 429} (2006) 307
  [arXiv:astro-ph/0603494].

\bibitem{Hannestad:2006zg}
  S.~Hannestad,
  Ann.\ Rev.\ Nucl.\ Part.\ Sci.\  {\bf 56} (2006) 137
  [arXiv:hep-ph/0602058].

\bibitem{Hannestad:2003ye}
  S.~Hannestad and G.~Raffelt,
  JCAP {\bf 0404} (2004) 008
  [arXiv:hep-ph/0312154].

\bibitem{Hannestad:2005df}
  S.~Hannestad, A.~Mirizzi and G.~Raffelt,
  JCAP {\bf 0507} (2005) 002
  [arXiv:hep-ph/0504059].

\bibitem{Melchiorri:2007cd}
  A.~Melchiorri, O.~Mena and A.~Slosar,
  Phys.\ Rev.\  D {\bf 76}, 041303 (2007)
  [arXiv:0705.2695].

\bibitem{Hannestad:2007dd}
  S.~Hannestad, A.~Mirizzi, G.~G.~Raffelt and Y.~Y.~Y.~Wong,
  JCAP {\bf 0708}, 015 (2007)
  [arXiv:0706.4198].

\bibitem{Hannestad:2008js}
  S.~Hannestad, A.~Mirizzi, G.~G.~Raffelt and Y.~Y.~Y.~Wong,
  JCAP {\bf 0804}, 019 (2008)
  [arXiv:0803.1585].

\bibitem{Andriamonje:2007ew}
  S.~Andriamonje {\it et al.}  [CAST Collaboration],
  JCAP {\bf 0704}, 010 (2007)
  [arXiv:hep-ex/0702006].

\bibitem{Inoue:2008zp}
  Y.~Inoue, Y.~Akimoto, R.~Ohta, T.~Mizumoto, A.~Yamamoto
  and M.~Minowa,
  arXiv:0806.2230 [astro-ph].

\bibitem{Peccei:2006as}
  R.~D.~Peccei,
  Lect.\ Notes Phys.\  {\bf 741} (2008) 3
  [arXiv:hep-ph/0607268].

\bibitem{Yao:2006px}
  W.~M.~Yao {\it et al.}  [Particle Data Group],
  J.\ Phys.\ G {\bf 33} (2006) 1.

\bibitem{Raffelt:2006cw}
  G.~G.~Raffelt,
  Lect.\ Notes Phys.\  {\bf 741} (2008) 51
  [arXiv:hep-ph/0611350].

\bibitem{Sikivie:2006ni}
  P.~Sikivie,
  Lect.\ Notes Phys.\  {\bf 741} (2008) 19
  [arXiv:astro-ph/0610440].

\bibitem{Asztalos:2006kz}
  S.~J.~Asztalos, {\it et al.},
  Ann.\ Rev.\ Nucl.\ Part.\ Sci.\  {\bf 56} (2006) 293.

\bibitem{Chang:1993gm}
  S.~Chang and K.~Choi,
  Phys.\ Lett.\ B {\bf 316} (1993) 51
  [arXiv:hep-ph/9306216].

\bibitem{Masso:2002np}
  E.~Mass\'o, F.~Rota and G.~Zsembinszki,
  Phys.\ Rev.\ D {\bf 66} (2002) 023004
  [arXiv:hep-ph/0203221].

\bibitem{Hinshaw:2008kr}
  G.~Hinshaw {\it et al.},
  arXiv:0803.0732 [astro-ph].

\bibitem{Nolta:2008ih}
  M.~R.~Nolta {\it et al.},
  arXiv:0803.0593 [astro-ph].

\bibitem{lambda}
  Legacy Archive for CMB Data Analysis
  (LAMBDA), http://lambda.gsfc.nasa.gov

\bibitem{Dunkley:2008ie}
  J.~Dunkley {\it et al.}  [WMAP Collaboration],
  arXiv:0803.0586 [astro-ph].

\bibitem{Komatsu:2008hk}
  E.~Komatsu {\it et al.},
  arXiv:0803.0547 [astro-ph].

\bibitem{Percival:2006gt}
  W.~J.~Percival {\it et al.},
  Astrophys.\ J.\ {\bf 657} (2007) 645
  [arXiv:astro-ph/0608636].

\bibitem{Tegmark:2006az}
  M.~Tegmark {\it et al.},
  Phys.\ Rev.\  D {\bf 74} (2006) 123507
  [arXiv:astro-ph/0608632].

\bibitem{Cole:2005sx}
  S.~Cole {\it et al.} [2dFGRS Coll.],
  Mon.\ Not.\ Roy.\ Astron.\ Soc.\  {\bf 362} (2005) 505
  [arXiv:astro-ph/0501174].

\bibitem{Eisenstein2005}
  D.~J.~Eisenstein {\it et al.} [SDSS Collaboration],
  Astrophys.\ J.\  {\bf 633} (2005) 560
  [arXiv:astro-ph/0501171];
  see also
  http://cmb.as.arizona.edu/$\sim$eisenste/acousticpeak

\bibitem{Davis:2007na}
  T.~M.~Davis {\it et al.},
  Astrophys.\ J.\  {\bf 666} (2007) 716
  [arXiv:astro-ph/0701510].

\bibitem{Lewis:2002ah}
  A.~Lewis and S.~Bridle,
  Phys.\ Rev.\ D {\bf 66} (2002) 103511
  [arXiv:astro-ph/0205436]

\bibitem{cosmomc}
  A.~Lewis, Homepage, {\tt http://cosmologist.info}


\end{thebibliography}
\end{document}